\documentclass{aa}  

\usepackage{mathptmx}
\usepackage{txfonts}
\usepackage[T1]{fontenc}
\usepackage{ae,aecompl}
\usepackage{xcolor}  

\usepackage{graphicx}	
\usepackage{mathtools,mathrsfs}
\usepackage{natbib}
\usepackage{hyperref}
\usepackage{ulem}
\usepackage{caption}
\usepackage{subcaption}
\usepackage{verbatim}
\usepackage{comment}
\newcommand{\punit}{\rm {g cm^{-1} s^{-2}}}
\newcommand{\gaea}{\textsc{GAEA}\xspace}
\newcommand{\tng}{\textsc{TNG}\xspace }
\newcommand{\eagle}{\textsc{EAGLE}\xspace }
\newcommand{\magneticum}{\textsc{Magneticum Pathfinder}\xspace }
\newcommand{\tth}{\textsc{The Three Hundred}\xspace }

\begin{document}

   \title{The impact of ram pressure on cluster galaxies, insights from \gaea and \tng }
   \titlerunning{RPS on cluster galaxies}
    \author{Lizhi Xie \inst{1} 
          \and
          Gabriella De Lucia\inst{2,3} 
          \and
          Matteo Fossati \inst{4,5}
          \and
          Fabio Fontanot \inst{2,3} 
          \and
          Michaela Hirschmann \inst{6,2}
          }
    \institute{Tianjin Normal University, Binshuixidao 393, Tianjin, China \email{xielizhi.1988@gmail.com}
             \and
             INAF – Astronomical Observatory of Trieste, via G.B. Tiepolo 11, I-34143 Trieste, Italy 
             \and
             IFPU - Institute for Fundamental Physics of the Universe, via Beirut 2, 34151, Trieste, Italy
             \and     
              Dipartimento di Fisica G. Occhialini, Universit\`a degli Studi di Milano Bicocca, Piazza della Scienza 3, 20126 Milano, Italy
            \and     
              INAF - Osservatorio Astronomico di Brera, via Brera 28, 20121 Milano, Italy
             \and 
             Institute for Physics, Laboratory for Galaxy Evolution and Spectral Modelling, Ecole Polytechnique Federale de Lausanne, Observatoire de Sauverny, Chemin Pegasi 51, 
CH-1290 Versoix, Switzerland
             }

   \date{Received xxx; accepted xxx}
 


\abstract{
Ram pressure stripping (RPS) has a non-negligible impact on the gas content of cluster galaxies. We use the semi-analytic model \gaea and the hydro-simulation \tng to investigate whether cluster galaxies suffer a strong RPS that is sufficient to remove a significant fraction of their gas during the first pericentric passage. We estimate that a ram pressure of $10^{-10.5}$, $10^{-12} $, $10^{-13.5} \punit$ can remove at most $90\%$, $50\%$, and $20\%$ of the cold gas reservoir from low-mass galaxies with $9<\log M_{\star}/{\rm M}_{\odot} <9.5$, assuming the gas can be stripped instantaneously. We then use this information to divide the phase space diagram into `strong', `moderate', `weak', and `no' RPS zones. 
By tracing the orbit of galaxies since $2.5R_{vir}$, we find in both \gaea and \tng that about half of the galaxies in Virgo-like halos ($\log M_h / M_{\odot} \sim 14 $) did not suffer strong RPS during the first pericentric passage. In Coma-like halos ($\log M_h / M_{\odot} \sim 15$), almost all galaxies have suffered strong RPS during the first pericentric passage, which can remove all gas from low-mass galaxies but is insufficient to significantly reduce the gas content of more massive galaxies. In general, results from \tng and \gaea are consistent, with the RPS being only slightly stronger in \tng than in GAEA. Our findings suggest that most cluster galaxies will maintain a notable fraction of their gas and continue forming stars after the first pericentric passage, except for those with low stellar mass ($\log M_{\star}/{\rm M}_{\odot} <9.5$) in very massive halos ($\log M_{h}/{\rm M}_{\odot} > 15$). 
}


\maketitle


\section{Introduction}
\label{sec:introduction}

Satellite galaxies, especially those in massive halos, are more quiescent, redder, and HI-deficient than central galaxies of similar mass \citep{weinmann2006, wetzel2012, poggianti1999, haynes1984, gavazzi2005, dressler1980}. The trend results from a combined effect of physical processes involving gravitational and hydro-dynamical interactions between satellite galaxies and their surrounding environment. The cutoff of gas accretion \citep[strangulation,][]{larson1980}, direct stripping of the hot gas \citep[starvation,][]{balogh2000}, tidal stripping of interstellar medium (ISM) \citep{merritt1983}, and ram pressure stripping (RPS) of the ISM \citep{gunn1972} are considered to be the relevant processes for environmental quenching. Albeit the efficiency of ISM stripping is lower than that of hot gas stripping \citep{mcCarthy2008, zhu2024}, RPS is considered a non-negligible environmental effect supported by various observational evidence, including asymmetries in the gas-phase morphology, stripping tails, and truncated gas disks \citep[][and references therein]{cortese2021PASA, boselli2022A&ARv}.

The impact of RPS increases with host halo mass. At fixed stellar mass, galaxies in cluster halos ($M_{h} > 10^{14} {\rm M}_{\odot}$) are generally HI-poorer than those in groups and in the field \citep{brown2017}. \citet{gavazzi2018} found a dominant fraction of late-type cluster galaxies showing one-sided stripping tails. The fraction is much lower in group halos \citep{Roberts2021}. 
In cluster halos, the ram pressure exerted on a satellite galaxy passing close to the halo centre (i.e. pericenter) can reach a pressure of $10^{-10} g/cm^2/s$ \citep{roediger2005}, which exceeds the binding energy of the inner gas disk for a Milky-Way like galaxy. \citet{Bruggen2008} suggested that more than half of cluster galaxies have experienced ram pressure stronger than $10^{-11} g/cm^2/s$, and would lose a large amount of gas. The dependence on host halo mass is supported by large-scale cosmological simulations \citep{marasco2016}, as well as by semi-analytic models \citep{xie2020}.

For galaxies falling onto cluster halos, RPS is onset beyond the virial radius \citep{bahe2013, marasco2016, ayromlow2019}. In fact, asymmetries in the gas morphology or in the distribution of young stars in galaxies (often, but not always, called jellyfish galaxies) have been detected in galaxies at the outskirts of cluster halos \citep{chung2007, wang2020, piraino-cerda2024}, and may appear beyond 2 times of the virial radius as found in the \tng simulation \citep{Zinger2024}. In the inner regions of clusters, galaxies are found to be either HI-deficient or to exhibit stripping tails that are aligned with the galaxies' orbit \citep{boselli2016, jaffe2016}. The HI deficiency increases towards the host halo centre, while the gas disk size decreases \citep{giovanelli1985, cortese2012,fossati2013,jaffe2016, reynolds2022}.

The first pericenter, where the galaxy experiences the peak ram pressure, is found to be crucial by many hydro-simulations, as most satellite galaxies lose their entire gas reservoir when approaching it \citep{Jung2018, lotz2019}. \citet{Wright2022} suggested that galaxies with a mass of $10^9 -10^{10} M_{\odot}$ can be quenched on the first orbit around massive halos with $M_h > 10^{15} M_{\odot}$, and can be quenched after the second pericenter in lower-mass halos. 
Tracing the journey of satellite galaxies in observation is impossible, but the phase space diagram, defined by the line-of-sight velocity and the projected distance to the cluster centre, has proven to be a useful diagnostic for the galaxy orbit and its infall time \citep{jaffe2015, rhee2017, pasquali2019}. By dividing the phase-space diagram into recent infall and virialised zones, \citet{jaffe2018} found that jellyfish galaxies are recently accreted and move on highly radial orbits. Other studies have confirmed that cluster galaxies that penetrate deeper into cluster halos exhibit more significant perturbations of their morphology \citep{vulcani2017, luber2022, biviano2024}.
Using an analytic model and an orbits library extracted from N-body simulations \citep{oman2016}, \citealt{oman2021} analysed the infall orbits of SDSS galaxies and concluded that galaxies with $10^{9.5}<M_{\star}<10^{11} {\rm M}_{\odot}$ in halos with $M_h > 10^{14} {\rm M}_{\odot}$ likely lose their entire gas reservoir around or before the first pericenter. 
\citet{wang2021} analysed the radial profile of the restoring force of resolved galaxies ($8.4< \log M_{\star}/{\rm M}_{\odot} <10.9$) and ram pressure distribution in the phase-space diagram of the Hydra cluster from the WALLABY survey\citep{Koribalski2020}. They estimated that RPS starts being effective at $\sim 1.25 R_{200}$, but that only a ram pressure of $10^{-10} \punit$, which requires a projected distance of $0.25 r_{200}$, can completely strip HI from most galaxies.  

In addition to the location of the pericenter, one should also consider the stripping time scale. 
In fact, gas stripping does not occur instantaneously. \citet{roediger2005} suggested a two-phase process: i) the outer disk gas is disturbed by ram pressure within a time scale of $10$~Myr, and ii) a continuous stripping that delays the mass loss for a time scale of 1~Gyr. The duration of the second phase depends on the strength of ram pressure and on the gravitational restoring pressure\citep{roediger2006, roediger2007}. When the ram pressure decreases as the galaxy moves away from the pericenter, the stripped tail can fall back to the galaxy\citep{vollmer2001}. \citet{jachym2009, koppen2018} suggested that ram pressure history is more important than the maximum pressure in determining the stripping efficiency.

Cosmological simulations can shed light on the impact of RPS across a broad range of stellar mass and environments. However, there is still no consensus among different studies, as galaxy quenching is a net effect influenced by many factors and is heavily dependent on the detailed treatment of various physical processes.  
In principle, cosmological hydro-dynamical simulations naturally account for hydro-dynamical processes like RPS, but are affected by limited resolution and internal feedback which could artificially enhance the impact of RPS \citep{bahe2017, Kulier2023}.
\citet{lotz2019} showed that in the \magneticum cosmological simulations RPS can remove the majority of the gas from cluster galaxies during their first pericenter passage. Consistent results were reported in studies based on \eagle \citep{Wright2022} and \tng \citep{rohr2023}. This very efficient gas stripping results in most satellite galaxies close to the cluster centres being completely depleted of HI in \tng and \eagle \citep{diemer2019, chen2024}. Similar findings have also been reported in \tth \citep{mostoghiu2021}. Moreover, \eagle, \tng, and \magneticum over-predicts the quenched fractions of low-mass galaxies in galaxy clusters \citep{bahe2017, lotz2019, donnari2021}. This trend is shared by other hydro-dynamical simulations at higher redshift \citep{kukstas2023}.  
RPS has also been considered in galaxy formation models based on a semi-analytic approach, with sometimes contradicting results about the efficiency of this process. For example, combining semi-analytic models and non-radiative hydrodynamic simulations, \citet{tecce2010} concluded that over 70 per cent of cluster galaxies within one virial radius have lost all their cold star-forming gas due to RPS.
More recent semi-analytic models \citet{cora2018, xie2020} have weakened the impact of RPS by considering the `shielding'  from hot and more diffuse gas associated with infalling satellites. Cluster galaxies in these models are primarily quenched by strangulation over a long timescale, whereas RPS has a minor effect, only diminishing the HI content of low-mass galaxies \citep{stevens2017}. As a consequence, most of the cluster galaxies with $M_{\star} > 10^{10} {\rm M}_{\odot}$ within $0.25 R_{200}$ still retain HI gas with a mass greater than $10^{8} {\rm M}_{\odot}$ \citep{chen2024}.

In this work, we investigate the strength and duration of RPS in cluster halos by tracing the orbital evolution of satellite galaxies to understand to what extent the first pericenter after infall represents a 
`dead end' for cluster satellites. We use both semi-analytic models and hydro-dynamic simulations and focus on the ram pressure derived from the orbital evolution of simulated galaxies. We do not discuss here the evolution of the gas content and star formation rates of satellite galaxies, as these properties require a complicated interpretation due to their sensitivity to progenitor bias and the combined effect of RPS and strangulation, which significantly depend on the physical models. The paper is organized as follows: we introduce the semi-analytic model \gaea and the \tng simulation in Section~\ref{sec:model}. In Section~\ref{sec:results}, we quantify the gravitational binding pressure of progenitor galaxies (Section~\ref{subsec:pressure}), derive fitting functions for the ram pressure distribution in the phase-space of $z=0$ cluster halos and their progenitors (Section~\ref{subsec:RPSzone}), trace the orbit of satellite galaxies and measure the duration of RPS (Section~\ref{subsec:dt}). Our conclusions are given in Section~\ref{sec:conclusion}.

\section{Models and methods}
\label{sec:model}

\subsection{GAEA semi-analtyic model}
\label{subsec:model}

In this work, we make use of the latest version of the GAlaxy Evolution and Assembly semi-analytic model \citep[][GAEA]{delucia2024}. The model originates from the one presented in \citet{delucia2007}, but has been further developed to include a non-instantaneous stellar recycling \citep{delucia2014}, an updated supernovae feedback model based in part on sophisticated hydro-dynamical simulations \citep{hirschmann2016}, a self-consistent partition of cold gas into HI and H$_2$ \citep{xie2017}, environmental processes \citep{xie2020}, as well as an improved treatment for accretion onto super-massive black holes and AGN feedback \citep{fontanot2020}. Model predictions are found to agree well with the observed stellar mass function for quenched and star-forming galaxies up to $z\sim 3-4$, quenched fractions up to $z\sim 3-4$, as well as other observational scaling relations at low redshift\citep{delucia2024, xie2024, fontanot2024}. 
Additionally, it is capable of reproducing the observed HI and H$_2$ fractions of central and satellite galaxies separately at $z\sim 0$ \citep{xie2020}, making it an excellent tool for our analysis. Below, we briefly summarize our treatment of environmental processes. For full details on our modeling, we refer to \citet{xie2020}.

In our model, galaxies are classified as central and satellite galaxies. Central galaxies are located at the centre of the dark matter halo potential, while satellite galaxies are those residing within the same friends-of-friends (FOF) group \citep{davis1985}. 
Satellite galaxies are subject to environmental effects, including strangulation/starvation, and direct stripping of their interstellar medium (ISM). These two processes affect both the hot and the cold gas. Here `hot gas' refers to the diffuse gas associated with the satellite, while `cold gas' refers to the neutral and ionised gas distributed within the gas disk.
Once a galaxy becomes a satellite,  gas accretion from the circum galactic medium onto the satellite's hot gas is suppressed. The hot gas cannot be replenished except by the re-incorporated gas that was previously ejected by SN and AGN feedback. The hot gas, as well as the ejected reservoir, can also be removed by ram-pressure stripping (RPS) and tidal stripping. 
The cold gas can be reduced by RPS if 1) it is located at radii larger than the stripping radius of the hot gas, and 2) the ram pressure is stronger than the gravitational binding energy. 

The model is run on the Millennium Simulation \citep{Springel2005} with a box size of $685$ Mpc based on a WMAP 1-yr cosmology \citep{spergel2003} with $\Omega_m = 0.25$, $\sigma_b = 0.045$, $\sigma_8 = 0.9$, and $h = 0.73$. We select galaxies in stellar mass bins $\log M_{\star}/M_{\odot}\sim$ [9, 9.5], [10, 10.5], and [11, 11.5] in the following studies. We also select galaxies in two mass bins of cluster halos, including the Virgo-like halos with $\log M_h/{\rm M}_{\odot} \sim [14, 14.5]$ and Coma-like halos with $\log M_h/{\rm M}_{\odot} \sim [15, 15.5]$. 

\subsection{The \tng simulation}

In this study, we also use results from illustris\tng \citep{weinberger2017, springel2018, Nelson2018, Naiman2018, Marinacci2018, Pillepich2018a, pillepich2018b} for consistency check. The \tng project is a suite of cosmological magneto-hydro-dynamical simulations adopting a Planck cosmology \citep{planck2015} with $\Omega_m = 0.3089$, $\Omega_b = 0.0486$, $\sigma_8 = 0.8159$, and $h = 0.6774$. We used \tng-100 which has a side length of the simulation box of approximately 100 Mpc. The gas mass resolution is $1.4 \times 10^6 {\rm M}_{\odot}$. 

We use both group catalogues and cell data from \tng. The group catalogues provide information about dark matter halos (and subhalos) identified using the FOF algorithm as well as galaxy properties. We use the mass of stars within twice the stellar half-radius as galaxy stellar mass.
The position and velocity of a galaxy are the position of the most bound particle and the sum of the mass-weighted velocities of all particles in a subhalo, respectively. We consider all star-forming gas cells and gas cells with an effective temperature lower than $10^{4.5}$~K as cold gas.  

From the \tng simulation, we also select galaxies in the three stellar mass bins as for \gaea. Since we used a smaller box for the \tng, we only select the Virgo-like halos  from the \tng simulation.

\subsection{Orbital parameters of satellite galaxies}
\label{subsec:orbitpara}

Throughout this work, we study galaxies in cluster halos with mass $M_{200} > 10^{14} {\rm M}_{\odot}$ at $z=0$. We normalize the three-dimensional cluster centric distance $r_d(3D)$ and the relative velocity $V_{rel}$ between central and satellite galaxies by the virial radius $R_{vir,z}$ and virial velocity $V_{vir,z}$, respectively.  
The infall time is defined as the first time a satellite crosses 2.5$R_{vir,z}$. Below, the redshift of infall is referred to as $z_{2.5Rvir}$. We use this value rather than the virial radius to diminish the contribution from back-splashed and pre-processed galaxies\citep{bahe2013, oman2021}.

The first pericenter is defined as the minimum radius of the satellite's orbit after infall. In the following, the radius and time at pericenter are written as $R_{FP}$ and $T_{FP}$, respectively. The first apocenter corresponds to the maximum normalised distance to the halo centre after the first peri-center. The corresponding radius and time are indicated as $R_{AP}$ and $T_{AP}$. Below we refer to the journey from infall to the first apocenter" as the 'first pericenter passage'.
In order to resolve the pericenter and apocenter more accurately, we interpolate the radii and relative velocities of satellite galaxies with a time step of 0.02~Gyr using spline. An example of the interpolation is shown in the Appendix~\ref{sec:app_interp_orbit}.

In GAEA, a galaxy inherits the positions and velocities of its host halo/subhalo. For orphan galaxies whose subhalos can no longer be resolved, their trajectory follows that of the most bound particles of the last identified host subhalos. Therefore, the orbits of orphan galaxies should be considered with caution. 
However, orphan galaxies in \gaea represent only $\sim 22$ per cent of the total satellite sample at the first pericenter so they should not have a significant impact on our results. We also use the \tng simulation for comparison to understand how robust the results are.

\subsection{Calculation of pressures}
\label{subsec:pressure_calc}

\subsubsection{\gaea}

In \gaea, like in all semi-analytic models,  galaxies are not spatially resolved. We assume that both gaseous and stellar disks follow exponential density profiles. 
The gravitational binding pressure at a given radius $r$ is:
\begin{equation}
    P_{grav}(r) = 2 \pi G\Sigma_{gs} (<r) \Sigma_{g}(r),
    \label{eqn:gravity}
\end{equation}
where $\Sigma_{gs}$ is the surface density of the combined components of cold gas and stars within the radius $r$, and $\Sigma_{g}(r)$ is the surface density of cold gas within a ring at $r$.

The ram pressure suffered by each galaxy can be calculated using\citep{gunn1972}:
\begin{equation}
    P_{ram} = \rho_{\rm ICM} v^2
    \label{eqn: rp}
\end{equation}
where $\rho_{\rm ICM}$ and $v$ are the density of the local intra-cluster medium (ICM) and the relative velocity in the rest-frame of the ICM. For simplicity, we assume that the ICM density follows a singular isothermal sphere (SIS) distribution. The density profile can then be written as $\rho_{\rm ICM}(r) = \frac{M_{hot}}{4\pi R_{vir}r^2}$. Here $M_{hot}$ and  $R_{vir}$ represent the mass of the hot gas and the corresponding virial radius. 
We also assume that the relative velocity in the rest frame of the ICM is equal to the relative velocity with respect to the central galaxy. 
These assumptions are consistent with the ram pressure stripping model adopted in GAEA. It is worth stressing that we measure the ram pressure within 3$R_{vir,z}$ around the cluster halos, disregarding the existence of neighbour halos. This implies that our estimate represent a lower limit, as the presence of nearby over densities will lead to higher ram pressure estimates.

\subsubsection{\tng}

We use two different methods to calculate the gravitational binding pressure for galaxies from \tng. In the first case, we use equation~\ref{eqn:gravity} in a similar approach as for \gaea galaxies. We first identify the three axes of the distribution of cold gas cells; we then use the long and intermediate axes to define the cold gas disk. We project the gas cells and star particles on the cold gas disk and calculate the surface density of stars and gas (including hot and cold gas) within a radius $r$, and the surface density of cold gas within a ring at $r$. 
Alternatively, we use the method proposed in \citet{mcCarthy2008}:
\begin{equation}
    P_{grav}(r) = 2\frac{G M_{gs}(<r) \rho_g(r)}{r}
\end{equation}
Here, $M_gs(<r)$ is the total mass of gas and stars within the radius $r$, and $\rho_g(r)$ is the volume density of the cold gas at this radius. It is worth noting that gravity from central massive black holes and dark matter are not considered in either method. Therefore, the pressure estimates shown below should be considered as lower limits.

The ram pressure for each \tng galaxy is derived from equation~\ref{eqn: rp} using two different methods to compute the ICM density and the relative velocity. The first method corresponds to the method adopted for \gaea. The ICM density is assumed to have a SIS profile, and its mass is defined as the total mass of all gas cells in the FOF group minus the gas mass within the stellar half-mass radius of the central galaxy. The relative velocity is approximated by the velocity of the satellite galaxy in the rest frame of the central galaxy. 
Aside from this simplified procedure, we also use a more complicated method that accounts for the local ICM density and the relative velocity against local ICM based on gas cells. We select gas cells within a shell of $1.25<r/R_{sub}<2$ around each subhalo as the local background environment, where $R_{sub}$ is the radius corresponding to the maximum of the rotation curve. We choose this shell to avoid contamination from other close subhalos as suggested by \citet{ayromlow2019}. 
The ICM density is then defined as the density of ionized gas in this shell, and calculated using $\frac{\Sigma_i m_i (1-f_{n,i})}{V_{shell}}$. Here, $m_i$ is the mass of the gas cell $i$, and $f_{n,i}$ is the neutral gas fraction of this cell. $V_{shell}$ is the volume of the shell. The velocity of the ICM is $\frac{\Sigma_i m_i  (1-f_{n,i}) v_{i} }{\Sigma_i m_i (1-f_{n,i})}$, where $v_i$ is the velocity of the gas cell $i$.

\section{Results}
\label{sec:results}

In this section, we first quantify the gravitational binding pressure of galactic disks for central and satellite galaxies at given galaxy radii in section~\ref{subsec:Pgrav}. This analysis provides insights on the strength of the ram pressure required to perturb the gas disks. Then, in section~\ref{subsec:RPSzone}, we examine the distribution of the ram pressure within the phase space of cluster halos at different redshifts, dividing the phase space into four zones based on the ram pressure strength. Finally, in section~\ref{subsec:dt}, we analyse the orbit of satellite galaxies as they fall onto present-day cluster halos, and measure the time they spend in different RPS zones. 

\subsection{Gravitational binding pressure of disc galaxies}
\label{subsec:Pgrav}

\begin{figure*}
   \includegraphics[width=0.45\textwidth]{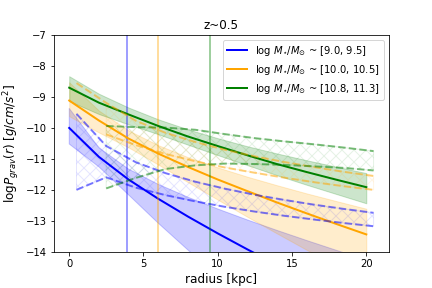}
   \includegraphics[width=0.45\textwidth]{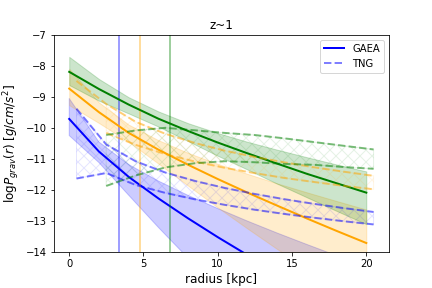}
   \caption{The gravitational pressure of gas discs as a function of physical-scale radius for central galaxies from \gaea at $z\sim 0.5$ (left) and $z\sim 1$ (right). The solid lines represent median values while the shadows show the 16th and 84th percentiles. Different colours represent different stellar mass bins as labelled. Solid vertical lines are the effective radius of cold gas disk for \gaea galaxies.
   The dashed lines show the corresponding measurements for \tng galaxies using two different methods (more details in the text).}
   \label{fig:Pgrav_r}
\end{figure*}

We select central galaxies with a gas fraction $f_g = \frac{M_{coldgas}}{M_{coldgas}+M_{\star}} > 0.1$ from the \gaea outputs at redshift $z\sim 0.5$ and $z\sim 1$, that are progenitors of cluster satellites that survive down to $z=0$. 
Figure~\ref{fig:Pgrav_r} shows the median pressure and the scatter for galaxies in three stellar mass bins. For any galaxy stellar mass, the restoring pressure decreases with increasing radius. At given radii, the pressure increases with increasing stellar mass. More massive galaxies are more resilient to RPS, as expected. The pressure also depends on the redshift. At $z\sim 1$, galaxies at a fixed stellar mass are more concentrated (smaller effective radii) than those at lower redshift $z\sim 0.5$. Consequently, the gravitational pressure at the centre of galaxies increases at higher redshift for a given stellar mass. To summarise, galaxies with larger stellar mass and at higher redshift are more resistant to ram pressure stripping. 

The restoring pressures of gas-rich central galaxies from \tng are shown as dashed lines in Figure~\ref{eqn:gravity}. 
For low-mass ($\log M_{\star}/M_{\odot}~[9,9.5]$) and intermediate-mass ($\log M_{\star}/M_{\odot}~[10,10.5]$) galaxies, the restoring pressure in the central regions of \tng galaxies is consistent with that of \gaea galaxies. At larger radii, the restoring pressure profile of \tng galaxies has a flatter slope and higher normalization compared to \gaea galaxies. These differences could be due to the higher temperature cut ($10^{4.5}$~K) for cold gas in \tng galaxies, and could be related to the different modelling of baryonic processes. The most massive \tng galaxies show decreasing restoring pressure in the central region, which is likely caused by the implementation of kinetic AGN feedback \citep{terrazas2020, shi2022}. 

We note that \gaea predictions are in good agreement with observed scaling relations between HI and H$_2$ fractions against stellar mass for star-forming galaxies at $z=0$, as well as with measurements of radii of stellar and star-forming gaseous disks up to $z\sim 2$\citep{xie2018, xie2020}. Therefore, we consider the gravitational pressure profile estimated from the density profile of stellar and gaseous components a reasonable representation of real galaxies. As a comparison, \citet{wang2021} measured the radial profile of the restoring pressure from the HI data of a few well-resolved galaxies around the Hydra cluster and found that the non-RPS galaxies with $M_{\star} \sim 10^{10} {\rm M}_{\odot}$ have restoring pressure between $10^{-11} \punit$ and $10^{-10} \punit$ within the r brand radius $r_{90}$. 
\citet{boselli2022A&ARv} found the gravitational pressure within the effective radius ranges between $10^{-13} \punit$ to $3\times 10^{-11} \punit$ for galaxies in nearby clusters. 
Our results are generally comparable to, if not slightly higher than, the observational measurements in the local universe.

\begin{figure*}
   \includegraphics[width=0.45\textwidth]{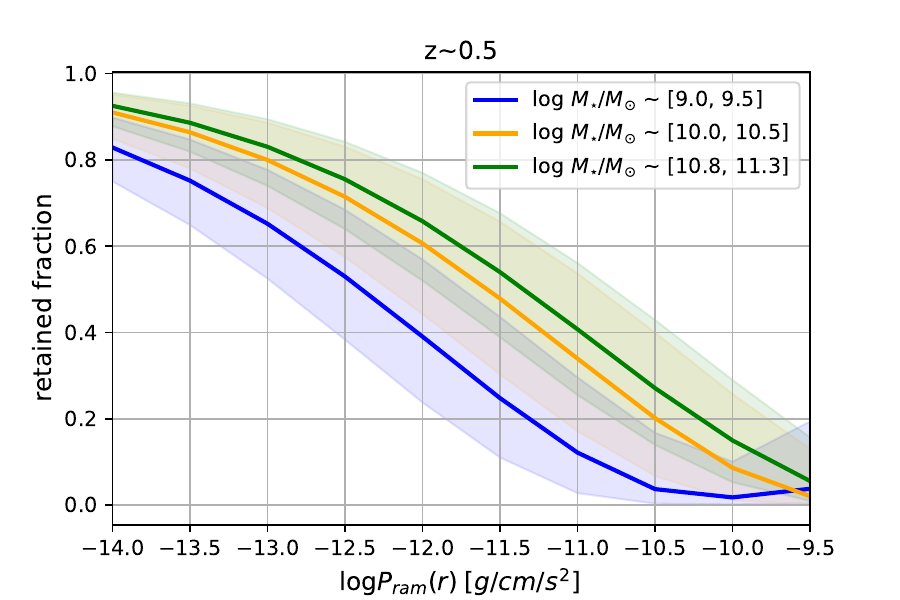}
   \includegraphics[width=0.45\textwidth]{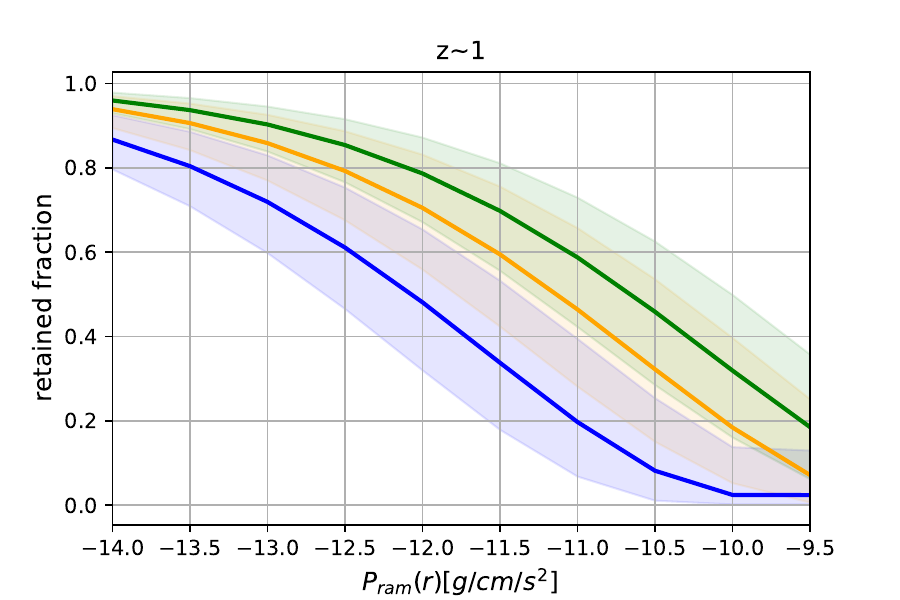}
   \caption{The retained gas fraction as a function of ram pressure. Colours and lines are the same as in Figure~\ref{fig:Pgrav_r}.}
   \label{fig:Pgrav_f}
\end{figure*}

Then we quantify the fraction of gas that can be retained, at a specific ram pressure strength, by simply assuming that gas at a given radius would be stripped off instantaneously when the ram pressure exceeds the gravitational binding pressure. Our estimated retained fractions are actually lower limits, because stripped gas takes some time to leave the galaxy. We only use \gaea for this analysis, as the profiles of \tng galaxies are not well approximated by exponentials, and a detailed calculation would require analysis on the gas parcels distribution as in \citet{Wright2022, rohr2023}, 
which is beyond the scope of this work. However, based on the profiles discussed above, we expect larger retained fractions for \tng for given ram pressure under the same assumption.

The results are shown in Figure~\ref{fig:Pgrav_f}. As expected, the retained fraction increases with increasing stellar mass and redshift. 
Specifically, a ram pressure above $10^{-10.5} \punit$ has a strong impact on low-mass galaxies by removing more than $90\%$ of the gas at $z\sim 0.5$ and $z\sim 1$. The effect is also significant for more massive galaxies, with only about $20-30\%$ of gas retained. A ram pressure of $10^{-12}  \punit$ can cause moderate stripping, leaving $40 \%$ and $60\%$ of the gas in low-mass and more massive galaxies at $z\sim 0.5$. 
The retained fraction increases to $\sim 50\%$ and $\sim 70\%$ at $z\sim 1$.
A ram pressure weaker than $10^{-13.5} \punit$ tends to have a negligible impact on both low-mass and more massive galaxies. 
Based on our analysis, we refer to ram pressure as `strong',  `moderate', `weak', and `none' using the following conditions: 
\begin{align*}  
  & \text{strong} \quad \log P_{ram} > -10.5, \\
  & \text{moderate} \quad -12 < \log P_{ram} < -10.5, \\
  & \text{weak} \quad -13.5 < \log P_{ram} < -12, \\
  & \text{none} \quad \log P_{ram} < -13.5.  
  \label{eqn:separation}
\end{align*}

\subsection{Ram pressure distribution in phase space}
\label{subsec:RPSzone}

In this section, we analyse the distribution of ram pressure in the phase space of cluster halos, to establish where galaxies are expected to be significantly affected by ram pressure stripping. We select cluster halos from both \gaea and \tng at $z=0$, and divide the sample by their host halo mass: $14 < \log M_{200}/{\rm M}_{\odot} < 14.5 $ (we refer to these as Virgo-like halos) and $15 < \log M_{200}/{\rm M}_{\odot} < 15.5 $ (Coma-like halos in the following). We note that \tng-100 does not include Coma-like halos.
Since halos keep evolving while satellite galaxies fall in, the progenitor halos at the moment of pericenter and infall time are less massive than the $z=0$ cluster halos we selected. In particular, we find that the progenitor halos have a median mass of $10^{13.8} {\rm M}_{\odot}$, and $10^{14.6} {\rm M}_{\odot}$ at $z\sim 1$ for Virgo-like and Coma-like halos, respectively.
To track the evolving environment of satellite galaxies, we trace the main progenitor of each halo back to higher redshift \footnote{We used the SubLink merger tree\citep{Rodriguez-Gomez2015} for \tng and LHaloTree \citep{Springel2005} for GAEA.}, and analyse the ram pressure distribution in their phase space.

For the \gaea clusters, we select all galaxies with $M_{\star} > 10^8 {\rm M}_{\odot}$ within 3 times $R_{vir}$ around each progenitor halo, and calculate the ram pressure based on their location in phase space. 

\begin{figure*}
    \includegraphics[width=0.5\linewidth]{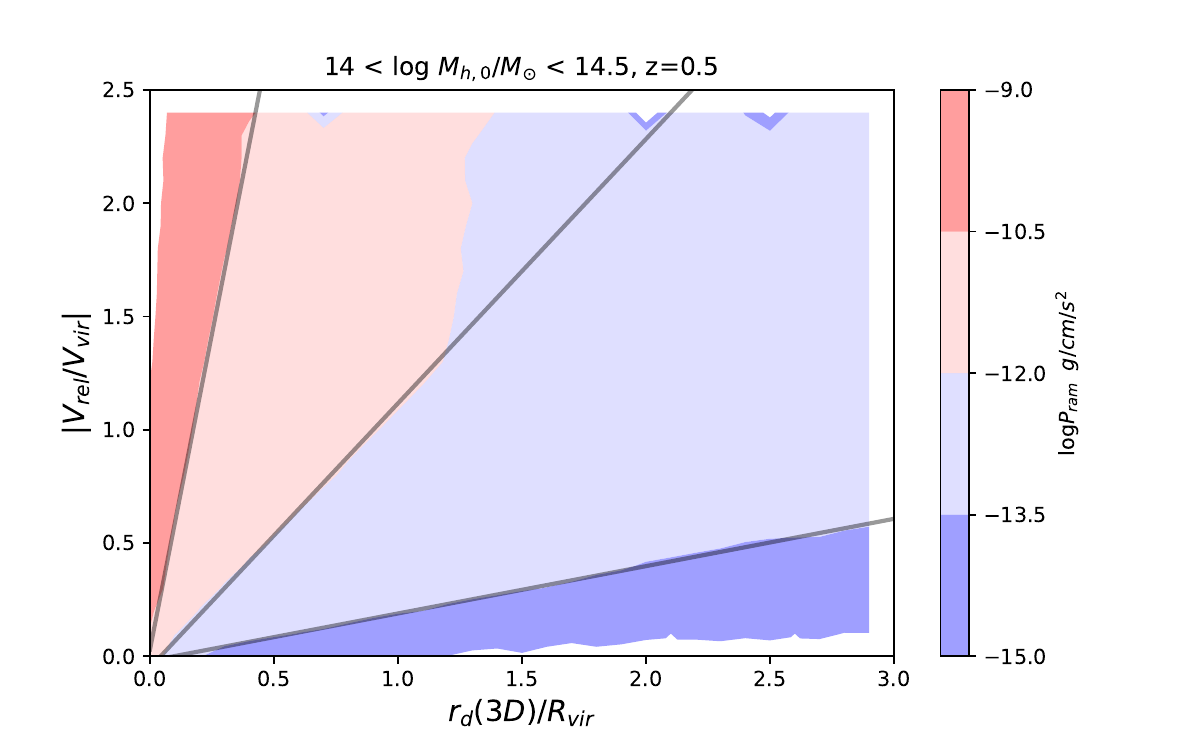}
    \includegraphics[width=0.5\linewidth]{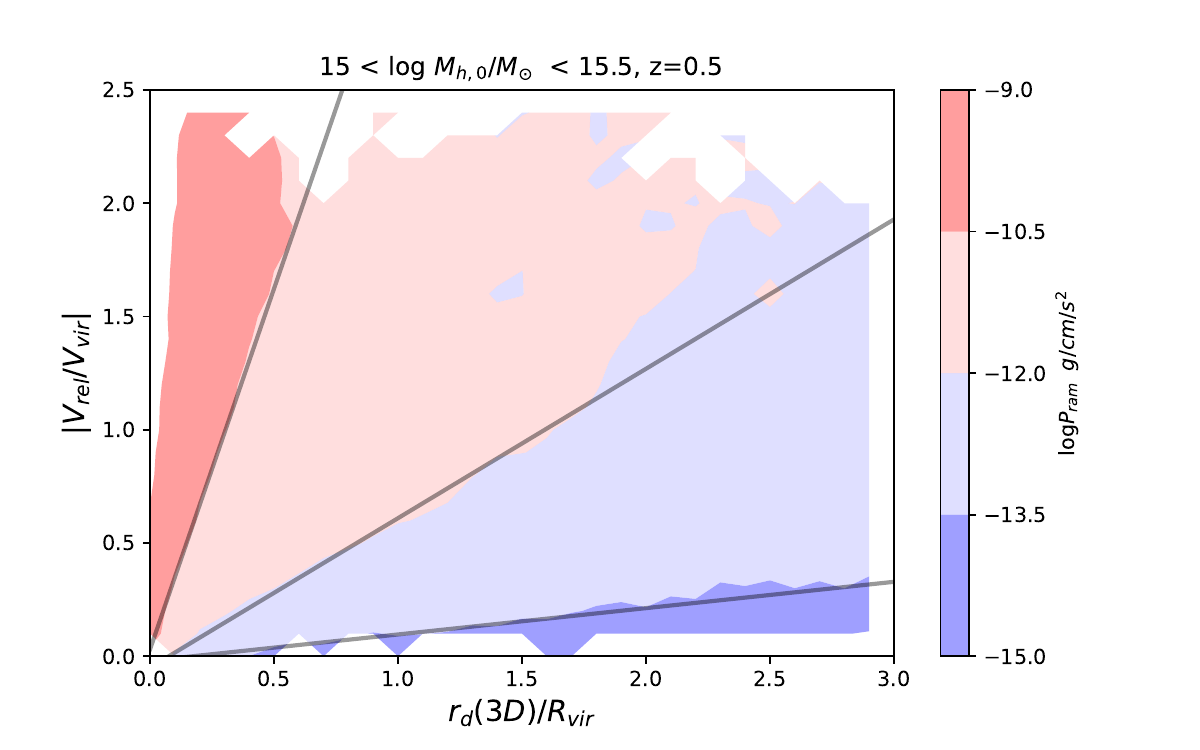}
    \includegraphics[width=0.5\linewidth]{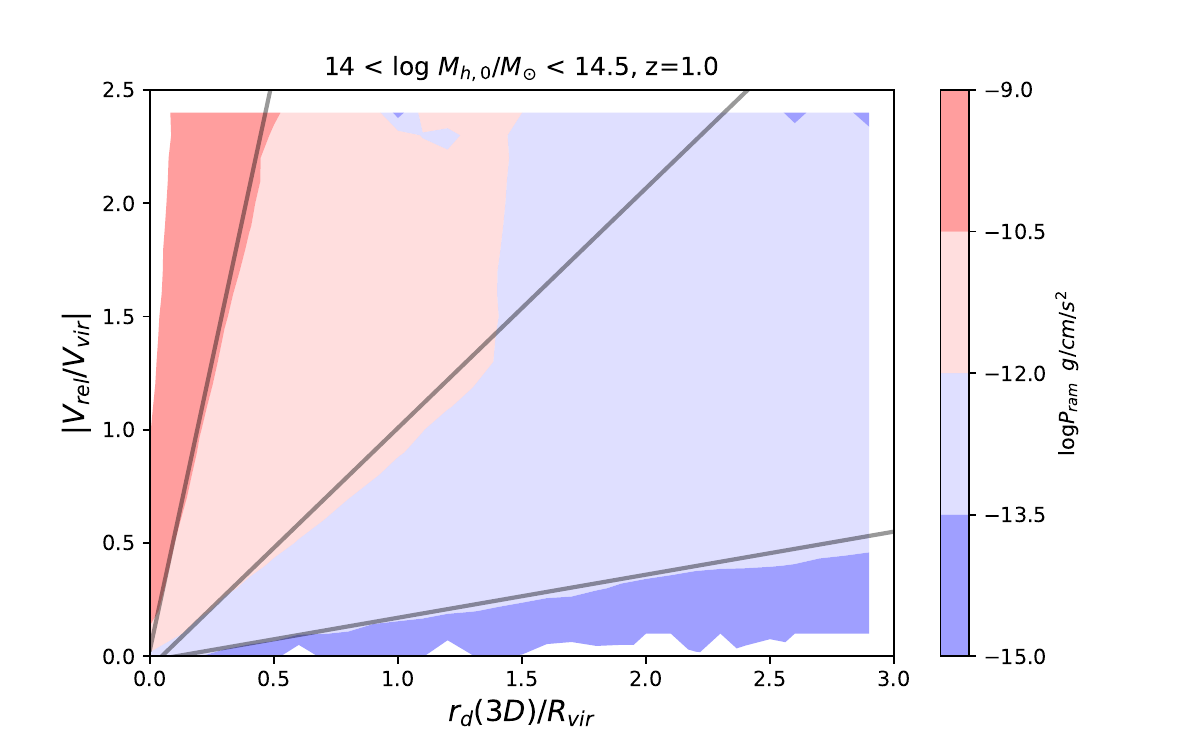}   
    \includegraphics[width=0.5\linewidth]{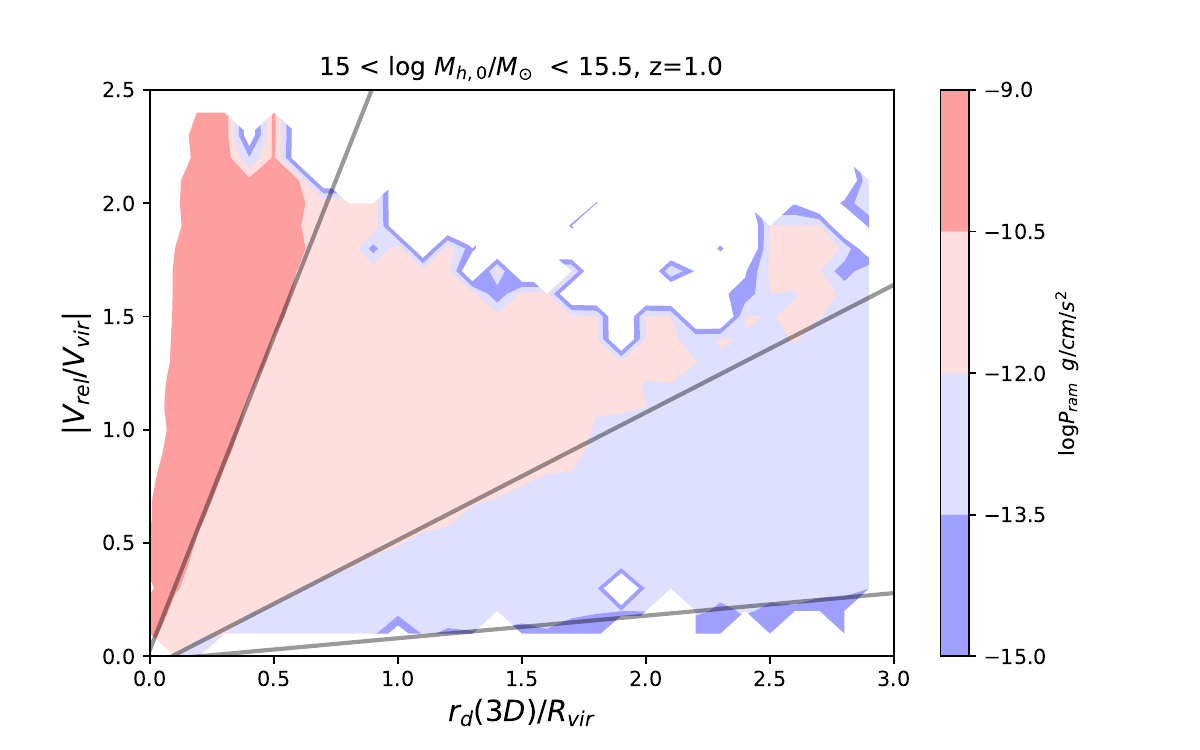}
    \caption{The median ram-pressure distributions around progenitors of the $z=0$ cluster halos from \gaea. The redshift and halo mass range are labelled above each panel. The x and y axes are 3D radius and velocity normalized by the virial radius and virial velocity. Different colors mark the four RPS zones defined in section~\ref{subsec:Pgrav}. Grey lines separating different zones are fits obtained as a function of redshift and host halo mass, defined by Equation:~\ref{eqn:fitting_sep}. 
    }
    \label{fig:rpzone}
\end{figure*}

Figure~\ref{fig:rpzone} shows the median ram pressure distribution in the phase space of cluster halos at $z=0$ from \gaea, and their progenitors at different redshifts. We divide the phase space into four zones by the characteristic ram pressure strength defined in the previous section. The separations between different zones are almost straight lines.  
We fit the separating lines and plot them as grey lines in Figure~\ref{fig:rpzone}:
\begin{align}
    & y = (-1.02 z -2.73\, \log M_{h,0} + 45)\, x+ 0.03, \\
    & y = (-0.23\, z -0.58\, \log M_{h,0} + 9.55)\,x  -0.05, \\
    & y = (-0.04\, z -0.106\, \log M_{h,0} + 1.74)\, x-0.02 .
  \label{eqn:fitting_sep}
\end{align}

Here $y$ and $x$ are normalized velocity $\frac{v}{V_{vir,z}}$ and cluster centric distance $\frac{r}{R_{vir,z}}$. $M_{h,0}$ is the final halo mass at $z=0$, and $z$ is the redshift.  
The ram pressure zones (RP zones) extend to larger normalised radii for more massive halos at higher redshift because of their higher concentration. 

\begin{figure*}
    \includegraphics[width=0.35\textwidth]{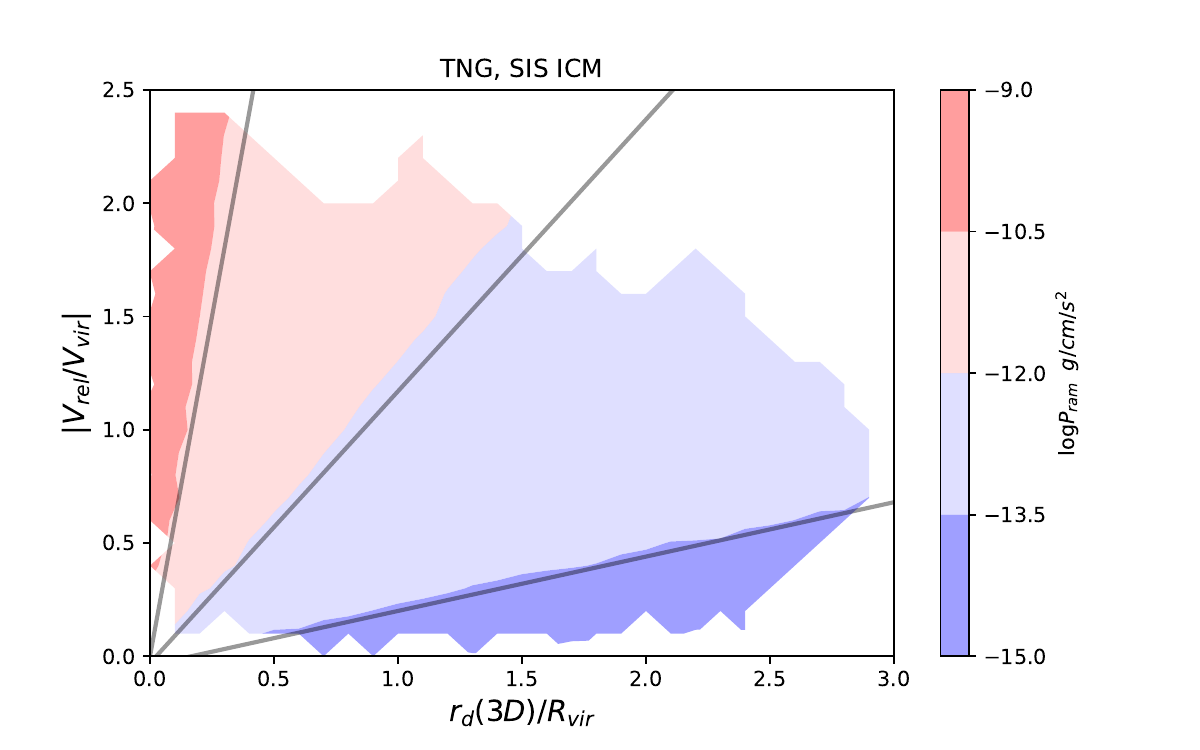}
    \includegraphics[width=0.35\textwidth]{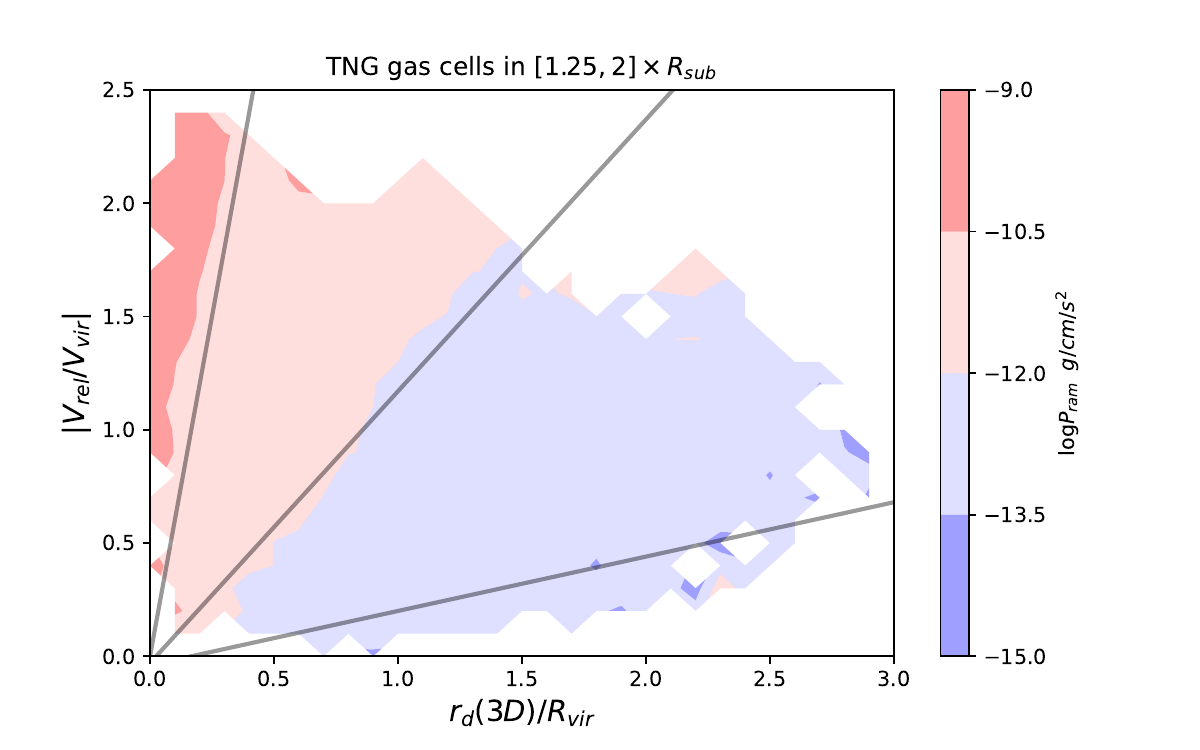}
    \includegraphics[width=0.26\textwidth]{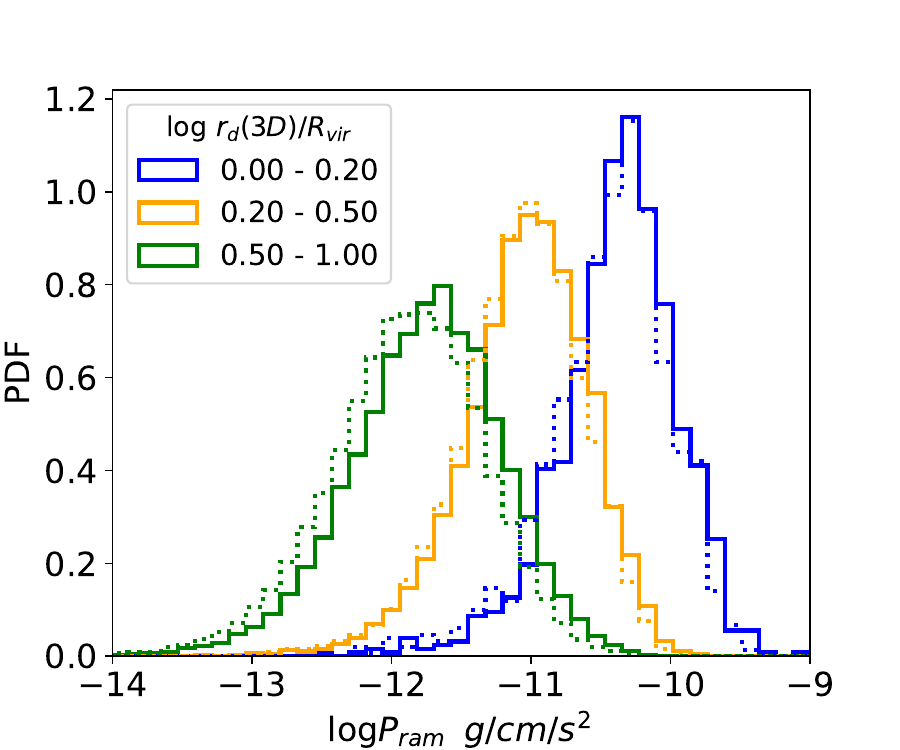}
    \caption{The RP distribution in phase-space around cluster halos from \tng at $z=0$. The ram pressure in left panel is measured using the simplified procedure as adopted for \gaea, while the ram pressure in the middle panel is derived from gas cells. 
    The right panel shows the distribution of ram pressure derived directly from gas cells and using the simplified procedure (solid and dotted histograms respectively).
    Different colors correspond to different cluster-centric distances, as indicated in the legend. 
    }
    \label{fig:TNG_rp_zone}
\end{figure*}

For the sake of simplicity, we have made several assumptions in the calculation of ram-pressure, e.g. we have assumed an isothermal distribution of the hot gas, and the relative velocity with respect to the central galaxy as the relative velocity in the rest-frame of the ICM. We have also ignored the orbital inclination. These are common simplifications adopted in semi-analytic models that lack internal spatial resolution. These assumptions, however, may have a profound impact on the measured distribution of ram pressure \citep{ayromlow2019}. In order to test their impact, we investigate the ram pressure zones of Virgo-like halos at $z=0$ from \tng using either the simplified procedure described above, or using a more complicated method analysing the gas cells. 
The results corresponding to the simplified method is shown in the left panel of Figure~\ref{fig:TNG_rp_zone}. The RP distribution calculated using gas cells information is shown in the middle panel of Fig.~\ref{fig:TNG_rp_zone}. These two algorithms provide consistent RP distributions, indicating that our results are robust. We further compare these two algorithms by plotting the probability distribution of ram pressure for galaxies selected in different cluster-centric distance bins in the right panel. For galaxies locate at larger radii, the ram pressure measured from gas cells tend to be stronger than that measured using simplified method. These two algorithms provide consistent measurements for galaxies in the inner part of cluster halos. 

For a comparison with data, \citet{boselli2022A&ARv} estimated the ram pressure exerted on galaxies that locate in Virgo cluster across a broad range of projected distance to halo center, and found that it ranges
between $10^{-12} \punit$ and $3\times 10^{-11} \punit$. 
\citet{jaffe2018} assumed a $\beta$ profile for cluster halos and estimated that the ram pressures at the centre ($0,1 R_{vir}$)  of cluster halos like Virgo/A85 are $\sim 10^{-11.5} /10^{-10.6} \punit$ for galaxies moving at a velocity of $1000 {\rm km/s}$, and $10^{-9.7} /10^{-10.5} \punit$ for a velocity of $3000 {\rm km/s}$.  These results are comparable to the ram pressure we measured for cluster halos in \gaea and \tng. \citet{wang2021} estimated the ram pressure distribution in the projected phase space of the Hydra cluster at $z\sim0$, and also found results that are consistent with our measurements.

\subsection{Ram pressure on satellite galaxies}
\label{subsec:dt}

\begin{figure*}
    \centering
    \includegraphics[width=0.95\textwidth]{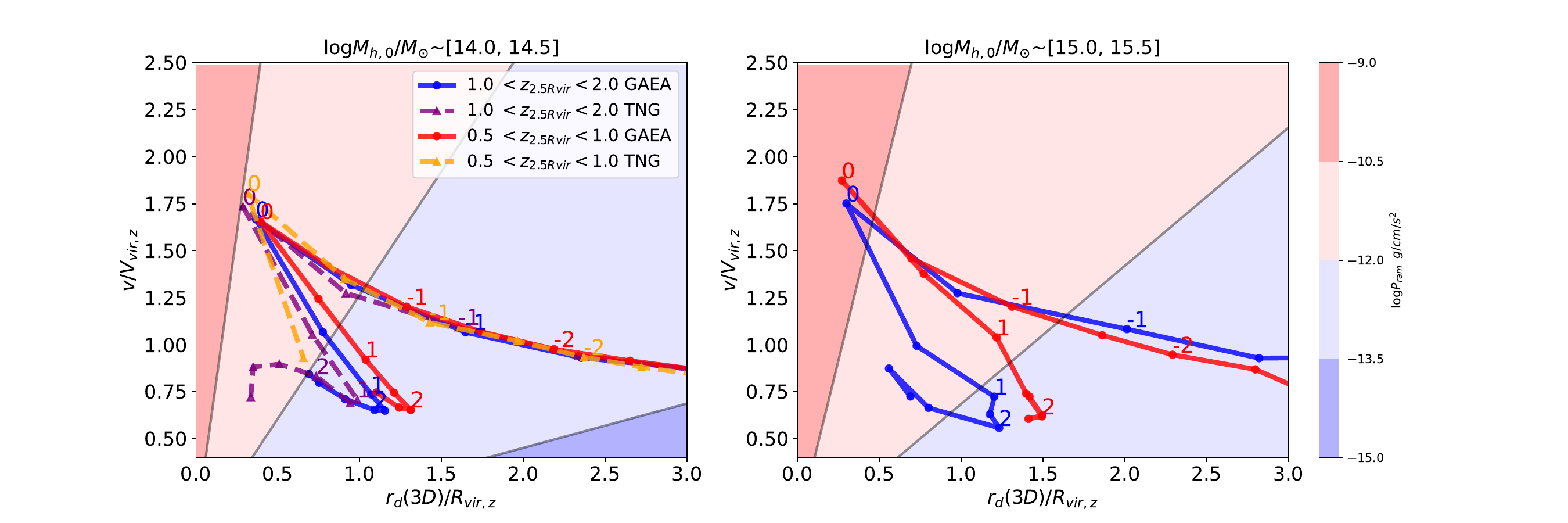}
    \caption{The orbit of galaxies ($9<\log M_{\star}/{\rm M}_{\odot} < 9.5$) falling onto the $z=0$ cluster halos at redshift $1<z<2$ and $0.5<z<1$. Red, pink, light blue, and blue shaded regions mark the strong, moderate, weak, and no RPS zones at $z\sim 0.5$. The left and right panels show the orbits of galaxies in different halo mass ranges. Different colors indicate different bins of infall time. Solid and dashed lines show results for \gaea and \tng galaxies. The numbers next to the orbits indicate the time $T-T_{\rm FP}$ in units of Gyr. Negative and positive values correspond to time before and after the pericenter. }
    \label{fig:orbit}
\end{figure*}

The maximum ram pressure at the pericenter gives an upper limit to the mass of stripped gas. Galaxies that penetrate deeper into the cluster halos, whose pericenters are located in the strong RPS zone, are expected to be more significantly affected by RPS. Those spending longer time in strong RPS zones are more likely to lose most of their gas. In this section, we analyse the orbit of cluster galaxies to quantify the RPS efficiency.

\begin{figure*}
    \centering
    \includegraphics[width=0.95\textwidth]{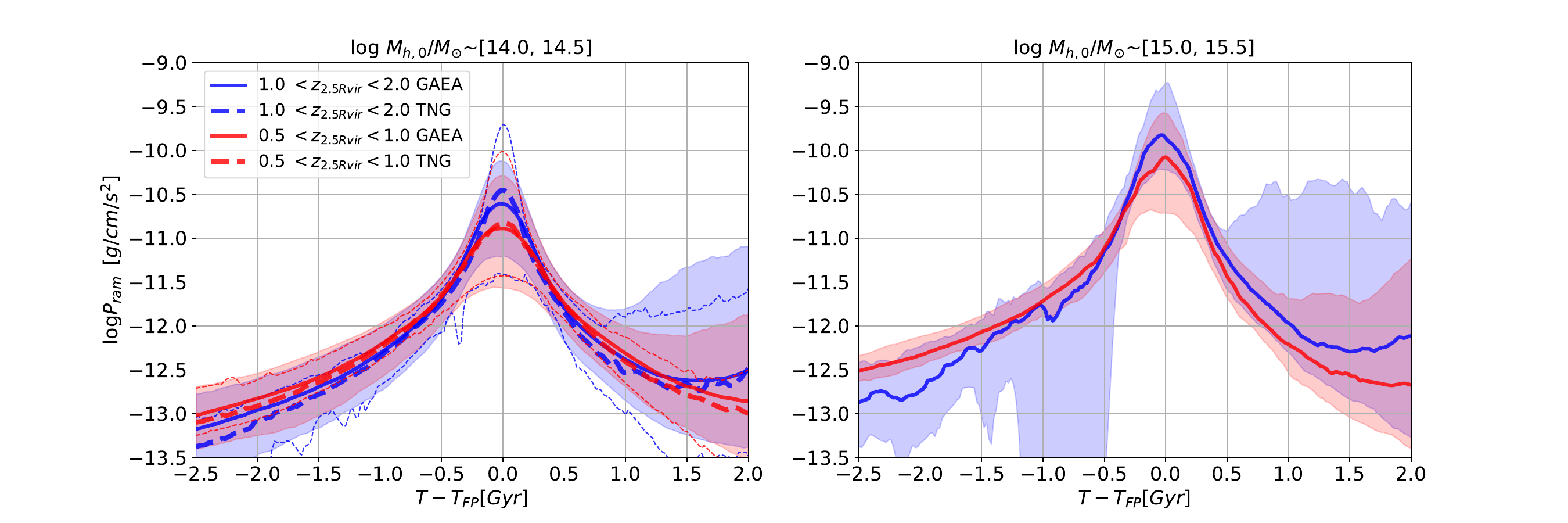}
    \caption{The median ram pressure on satellite galaxies with $9 < \log M_{\star} < 9.5$ as a function of time for \gaea (solid lines) and \tng  (dashed lines). Red and blue shades show the 16th and 84th percentiles for \gaea galaxies. Thin dashed lines show the corresponding scatter for \tng galaxies. Colors and line styles are the same as Fig.~\ref{fig:orbit}. }
    \label{fig:Pram_t}
\end{figure*}

From \gaea and \tng , we select satellite galaxies with $9 < \log M_{\star} < 9.5$ at $z=0$ that have already passed the first pericenter in cluster halos. We have checked that more massive galaxies have similar orbits as lower-mass ones, therefore we only show results for the low-mass galaxies. We split the galaxy sample by their infall time and host halo mass to study the dependence on these quantities. The orbits are rescaled with respect to the time when the galaxy reaches the first pericenter ($T-T_{\rm FP}$). The median orbits are shown in Figure~\ref{fig:orbit}. The strong, moderate, weak, and non-RPS zones, based on the fitting given in equations Eqn~\ref{eqn:fitting_sep} at $z\sim 0.5$, are marked as red, pink, light blue, and blue shaded regions. Fig~\ref{fig:Pram_t} shows the median ram pressure that satellite galaxies have experienced before and after the first pericenter.
We use the simplified method described in sec~\ref{subsec:RPSzone} to calculate ram pressure for both \gaea and \tng galaxies.  

Generally, galaxies in Virgo-like halos take about $2$~Gyr to reach the first pericenter after infall, and another $2$~Gyr to reach the first apocenter. At $3R_{vir,z}$, galaxies are already in the weak RPS zones (see Figure~\ref{fig:orbit}). The median ram pressure is approximately $10^{-13} \punit$ (see Fig~\ref{fig:Pram_t}), which could remove at most $30 \%$ of the gas from these galaxies (see Fig~\ref{fig:Pgrav_f}). These satellites spend over $1.5$~Gyr in the weak RPS zone, allowing removal of the stripped gas.
Satellite galaxies enter the moderate RPS zone at a radius of $\sim 1.25 R_{vir,z}$. At the pericenter, the ram pressure increases to as much as $\sim 10^{-10.5} \punit$. 
If the gas were removed instantaneously, the retained fraction of satellites would still be $10\%$ for low-mass galaxies, and more than $\sim 20\%$ for higher-mass galaxies on average. 
The retained fraction would be larger when considering other factors, i.e. a non-zero stripping time scale, reincorporation of the stripped gas, disc inclination and protection from the satellites' hot gas sphere. 
Therefore we expect satellite galaxies in Virgo-like halos to have lost a large fraction of their gas at the first pericenter, but not all of it, particularly the dense gas in the central regions that fuels star formation. Results from \tng are consistent with those based on \gaea, except for the fact that \tng galaxies exhibit a larger upper scatter of peak ram pressure.

Galaxies in Coma-like halos suffer stronger RPS, as expected. These galaxies spend more than $1$~Gyr in the moderate RPS zone before reaching the pericenter, leading to a complete removal of $50 \%$ of the gas from low-mass galaxies and $\sim 30\%$ from more massive galaxies. The maximum ram pressure can reach $\sim 10^{-10} \punit$, comparable to the restoring pressure at the centre of low-mass galaxies. Therefore, low-mass galaxies could lose all of their gas in Coma-like halos if stripping process occurs on short time-scales. Galaxies more massive than the Milky Way could retain $10-20 \%$ of their gas. These massive galaxies under strong ram pressure are potential candidates for jellyfish galaxies with long tails as detected in previous work \citep{luber2022}.

\begin{figure*}
    \centering
    \includegraphics[width=0.95\textwidth]{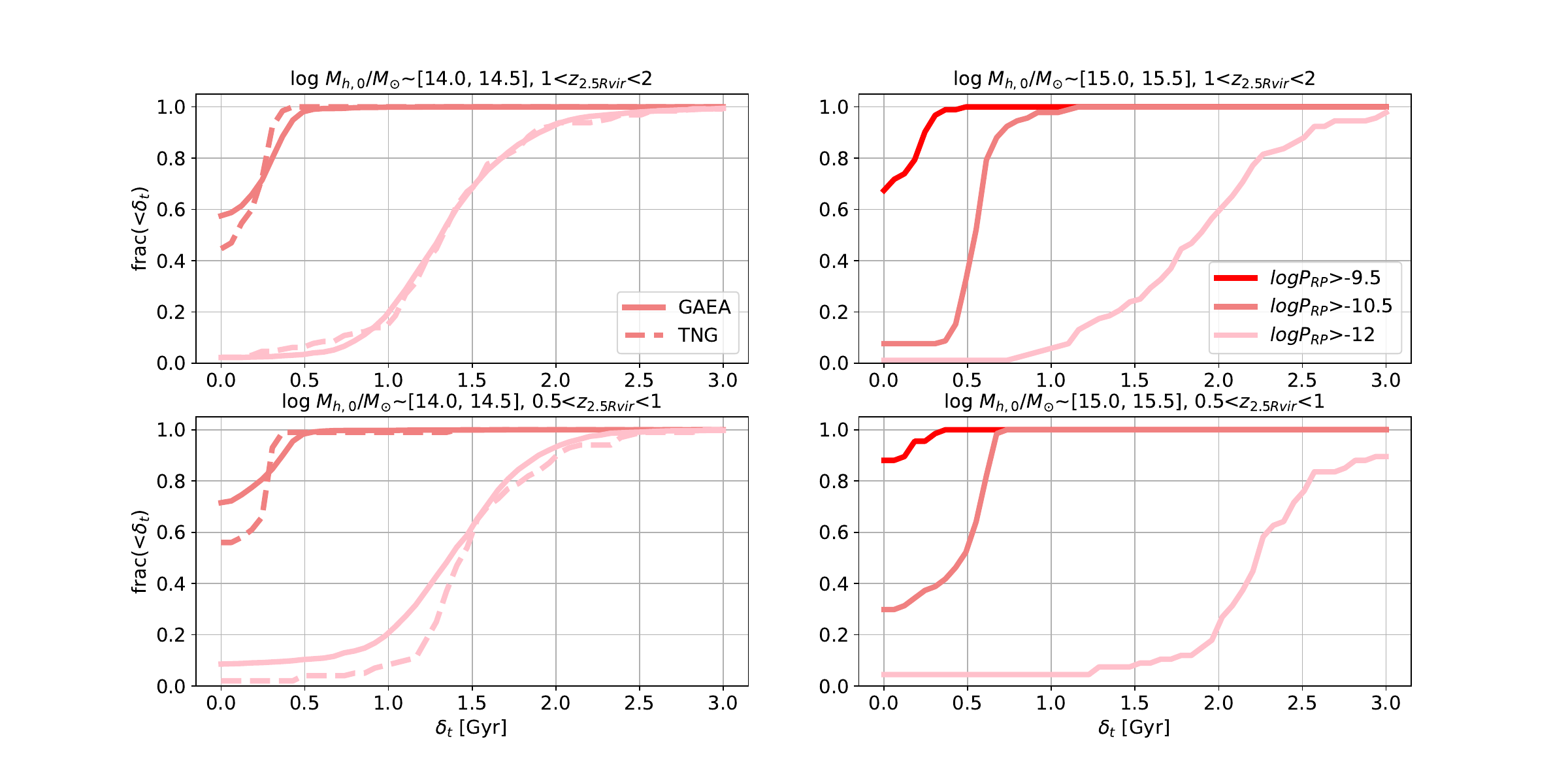}
    \caption{Fraction of galaxies that have spent less than a given time in various RPS zones during the first passage of their infall orbit. 
    We consider two bins of infall redshift and compute the cumulative times corresponding to three different lower limits for RPS, as indicated in the legend. 
    The y-axis is the fraction of galaxies that spend less than $\delta_t$~Gyrs in regions whose RPS value exceeds the one considered. We show the results for low-mass galaxies with $9 < \log M_{\star} < 9.5$ in this plot. The results do not depend significantly on stellar mass.
    }
    \label{fig:dt}
\end{figure*}

The maximum ram pressure at first pericenter has a large scatter as shown in Fig~\ref{fig:Pram_t}. To understand how much time galaxies spend in different ram pressure zones and the corresponding scatter, we sum the total time when the ram pressure exceeds a specific value during the first passage (from infall to apocenter) , 
and plot the cumulative distribution in Figure~\ref{fig:dt}. 
Galaxies that have not reached the first apocenter are excluded.

The left panel of Figure~\ref{fig:dt} shows results for galaxies in Virgo-like halos. More than $60\%$ of the galaxies have not experienced strong RP during their first passage in GAEA. The fraction is about $40\%$ for \tng galaxies. These galaxies are not expected to lose large fractions of their gas at the first passage due to ram pressure stripping. Those that have entered into the strong RPS zones stay less than $\sim 400$~Myrs. The duration is comparable to the time scale required for the stripped gas to leave the galaxy \citep{roediger2005}. A small fraction of galaxies in \gaea have not yet entered the moderate RPS zone. These galaxies can retain most of their gas. 
Most early infall galaxies spend over 1~Gyr in the moderate RPS zone. The long duration is sufficient to remove all the gas from the outer region. However, the remaining dense gas in the centre can still support star formation. 

In Coma-like halos, a small fraction, $\sim 10\%$ for early infall and $30\%$ for late infall, galaxies have not yet experienced strong RPS. Most galaxies spend over $500$~Myr in the strong ram pressure zone. All galaxies have experienced moderate RPS for more than $1$~Gyr during the first passage, likely leading to a significant depletion of gas for low-mass galaxies. For massive halos, we also plot duration for ram pressure above $10^{-9.5} \punit$. 
More massive galaxies have similar orbits as low-mass galaxies in our analysis. Most of the massive galaxies have not yet entered the extremely high ram pressure zone at the time when they reach the first apocenter.
Therefore we conclude that galaxies more massive than $10^{10} {\rm M}_{\odot}$ can retain at least $10-20\%$ gas after the first passage even in Coma-like halos.

\section{Summary}
\label{sec:conclusion}

Using the semi-analytic model \gaea and hydro-simulation simulation \tng, we investigate whether satellite galaxies can be significantly affected by RPS during the first pericenter passage as they fall on the cluster halos. 

We quantify the gravitational pressure for central galaxies in three stellar mass bins at different redshifts. The gravitational pressure increases with increasing stellar mass and redshift. At the centre and at the effective radii, it reaches values of approximately $10^{-10} \punit$ and $10^{-12} \punit$ for galaxies with $9<\log M_{\star}/{\rm M}_{\odot} < 9.5$ at $z\sim1$. These values increase to $10^{-9} \punit$ and $10^{-10.5} \punit$ for massive galaxies ($11<\log M_{\star}/{\rm M}_{\odot} <11.5$) at the same redshift.  We obtain generally consistent results for \tng and \gaea galaxies, with \tng central galaxies showing higher restoring pressure at large radii than \gaea.

We estimate the ram pressure exerted on galaxies within 3 times the virial radius of present-day clusters and use this information, combined with the estimates of gravitational pressure given above, to divide the phase space of into strong, moderate, weak, and no ram-pressure zones. We provide fits for the separation lines for ram pressure values of $10^{-10.5} \punit$, $10^{-12} \punit$, $10^{-13.5} \punit$ between these four zones  (see Equation~\ref{eqn:fitting_sep}). The ram pressure at given normalised location in phase-space is stronger in more massive halos and at higher redshift. Estimates based on \tng and \gaea are consistent, indicating our results are robust.

From the median orbits of satellite galaxies, we find that galaxies in Virgo-like halos have spent over $1$~Gyr in the moderate RPS zone from infall to the first apocenter. However, most of them avoid the strong RPS zone. Therefore, even low-mass galaxies with $9<\log M_{\star}/{\rm M}_{\odot} <9.5$ can retain at least $10\%$ of gas after the first passage around Virgo-like halos. In Coma-like halos, the RPS is sufficient to strip all gas from $>90\%$ of low-mass galaxies. The majority of more massive galaxies with $\log M_{\star}/{\rm M}_{\odot} > 10$ can still retain gas in their central regions after the first pericentric passage. The retained fraction would be higher considering other factors, i.e. finite stripping time scale, shielding effect from the surrounding atmosphere, etc. 

Our conclusion contradicts the results of previous studies based on cosmological hydro-dynamical simulations \citep{Jung2018, lotz2019, oman2021, Wright2022}, that find a complete/significant depletion of gas primarily due to RPS during the first pericenter passage for satellite galaxies in halos with $M_{h} > 10^{14} {\rm M}_{\odot}$. This significant reduction of the gas content eventually leads to their quenching. 
The gas depletion efficiency in these simulations may be overestimated as reflected by their over-predicted quenched fractions for the low-mass cluster galaxies compared to observations \citep{bahe2017, xie2020}. 
In this work, we find similar distribution of RPS in cluster halos and a higher gravitational binding pressure for central galaxies in \tng with respect to \gaea. Yet cluster galaxies in \tng are gas-poorer and more quiescent than those in \gaea. The RPS in \tng, as in many other hydro-dynamical simulations \citep[][for a review]{cortese2021PASA}, might be artificially enhanced due to resolution effects, mixing of ISM and ICM, and might be further enhanced by internal feedback processes which would reduce the binding pressure of cluster galaxies. Based on our analysis, we argue that longer timescales for gas depletion and quenching, as found in \gaea, are to be preferred and lead to model predictions that are in good agreement with the observed quenched fractions for low-mass galaxies. 

To conclude, our analysis supports a scenario in which galaxies in Virgo-like haloes are unlikely to lose their entire gas reservoir due to RPS during the first pericentric passage. The efficiency of stripping is more important for Coma-like halos, but only for low-mass galaxies. For our analysis, we only used the orbital evolution of model galaxies, which are independent of the modelling adopted for environmental effects. The consistency between two completely independent models and approaches considered in this study makes our results robust. We note that the ICM density is influenced by the feedback model adopted. We leave this question for a future study.

\section*{Acknowledgements}
LZX acknowledges support from the National Natural Science Foundation of China (grand number 12041302), the Ministry of Science and Technology of China (grant No. 2020SKA0110100). 

\section*{Data availability}

The data underlying this article will be shared on reasonable request to the corresponding author. 
An introduction to \gaea, a list of our recent work, as well as datafile containing published model predictions, can be found at \url{https://sites.google.com/inaf.it/gaea/home}. The \tng simulations are publicly available and accessible at \url{https://www.tng- project.org/data/}.



\bibliographystyle{aa}
\bibliography{references} 





\appendix

\section{Interpolation of the orbital evolution}
\label{sec:app_interp_orbit}

\begin{figure}
\includegraphics[width=0.4\textwidth]{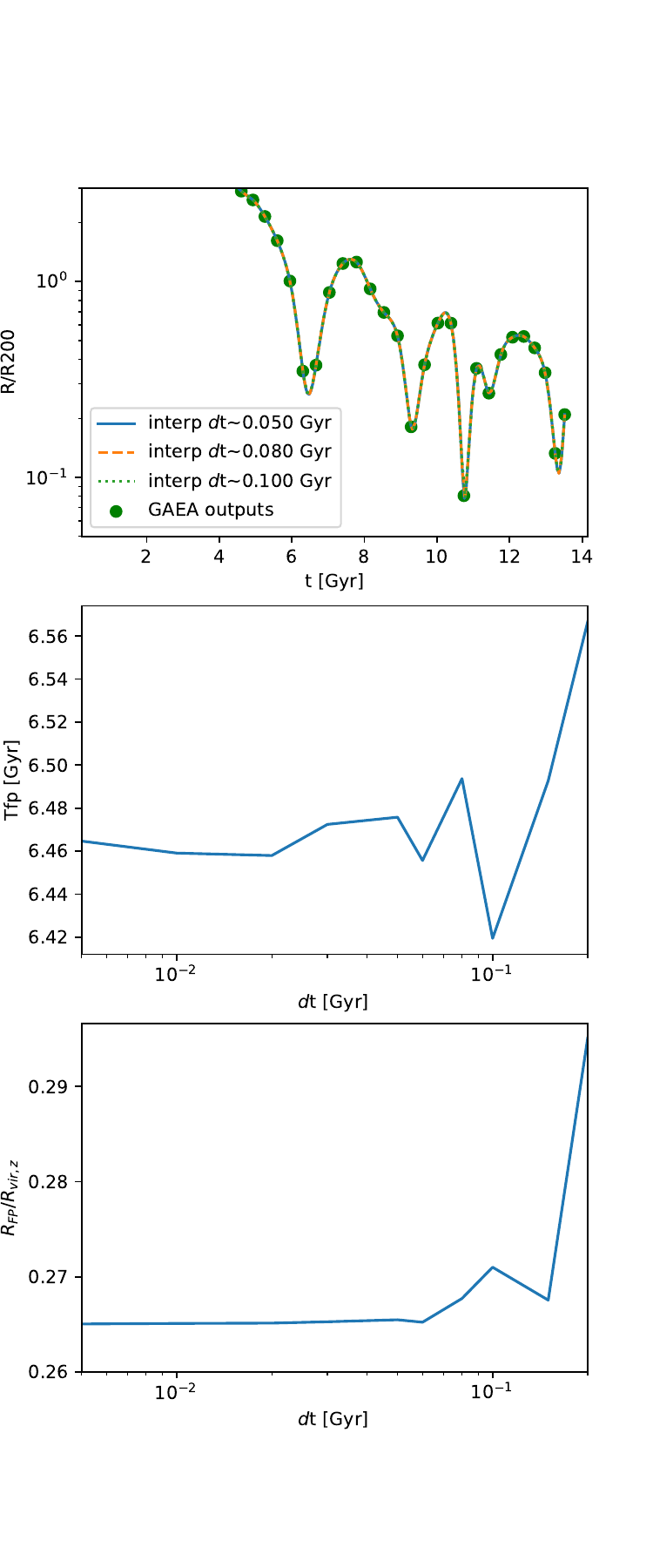}
\caption{An example of the interpolation of a satellite's orbit. The top panel shows the normalized distances to the cluster centre. Green points are model outputs, at the available snapshots of the simulation. Different lines show interpolations over different time steps. The middle and bottom panels are the $T_{FP}$ and $R_{FP}$ measured from the interpolated orbits as a function of the interpolation time-scale adopted. }

\label{fig:interp}
\end{figure}

Only a limited number of snapshots are available from the Millennium Simulation adopted in this study. To estimate more accurately the pericenter and apocenter of model galaxies, we interpolate the normalized cluster-centric distance at finer time steps using a spline function. Figure~\ref{fig:interp} shows the interpolated orbit and the interpolated orbital parameters of an example galaxy. We have tested time steps ranging from $1$~Myr to $0.2$~Gyr, and find that the measured $R_{FP}$ and $T_{FP}$ converge when the time-step is shorter than $0.02$~Gyr. 

\label{lastpage}
\end{document}